\documentclass[]{aa}
\usepackage[]{times}
\usepackage[]{graphics}
\usepackage{latexsym}
\begin{document}
\thesaurus{(13.09.1;11.09.1:M31;11.16.1;09.04.1;09.08.1)}

\title{The bright 175\,$\mu$m knots of the Andromeda Galaxy
\thanks{%
Based on observations with ISO, an ESA project with instruments funded
by ESA member States (especially the PI countries: France, Germany, the 
Netherlands, and the United Kingdom) and with the participation of ISAS and 
NASA.}}

\author{L. Schmidtobreick\inst{1,2} 
        \and M. Haas\inst{1} 
        \and D. Lemke\inst{1}}
\offprints{L. Schmidtobreick: linda@pd.astro.it}
\institute{Max-Planck-Institut f\"ur Astronomie, K\"onigstuhl 17, D--69117 Heidelberg, Germany
          \and Osservatorio Astronomico di Padova, Vicolo dell'Osservatorio 5, I--35122 Padova, Italy}
\date{Received 8 March 2000; accepted 7 September 2000}
\maketitle

\markboth{L. Schmidtobreick et al.: The 175\,$\mu$m knots in M\,31}
{L. Schmidtobreick et al.: The 175\,$\mu$m knots in M\,31}

\begin{abstract}

Discrete far-infrared (FIR) sources of M\,31 are identified in the ISO 175\,$\mu$m map and 
characterized via their FIR colours, luminosities and masses in order to reveal the 
nature of these knots. 
With our spatial resolution of 300\,pc at M\,31's distance, the FIR knots 
are clearly seen as extended objects with a mean size of about 800\,pc. 
Since this appears too large for a single dust cloud, the knots might 
represent several clouds in chance projection or giant cloud complexes.
 
The 175\,$\mu$m data point provides crucial information in addition to the 
IRAS 60 and 100\,$\mu$m data: 
At least two $\lambda$$^{\rm -2}$ modified Planckian curves with temperatures 
of 
about 40\,K and 15-21\,K 
are necessary to fit the spectral energy distributions (SEDs) of the knots. 
Though they show a continuous range of temperatures, we distinguish 
between three types of knots -- cold, medium, warm -- in order to recognize 
trends.   
Comparisons with radio and optical tracers show that -- statistically -- 
the cold knots can be identified well with CO and H\,I radio sources and thus 
might 
represent mainly molecular cloud complexes. 
The warm knots coincide with known H\,II regions and supernova remnants. 
The medium knots might contain a balanced mixture of molecular clouds and 
H\,II regions.
The cold knots have a considerable luminosity and their discovery 
raises the question of hidden star formation.

Though the optically dark dust lanes in M\,31 generally match the FIR ring, 
surprisingly we do not find a convincing coincidence of our knots with 
individual dark clouds, which might therefore show mainly foreground dust 
features. 

The ratio of FIR luminosity to dust mass, $L/M$, is used to measure the 
energy content of the dust. It can originate from both the interstellar 
radiation field and still embedded stars recently formed. 
The knots have a clear $L/M$ excess over the rest 
of M31, providing  evidence that they are powered by star formation 
in addition to the interstellar radiation field. 
Furthermore, the $L/M$ ratio of the warm knots is comparable to that of 
Galactic H\,II regions like M\,42 or NGC\,2024, 
while that of the cold knots still reaches values like in the average Orion 
complex. 
Thus both the warm and even the cold knots are interpreted as containing 
large cloud complexes with considerable ongoing star formation. 
%
%
\keywords{Infrared: galaxies --- Galaxies: individual: M31 -- photometry ---
ISM: dust, extinction -- H II regions}
\end{abstract}

\section{Introduction}
The ISO map of M\,31 at 175\,$\mu$m is dominated by a concentric ring
structure containing numerous individual knots
(Haas et al. \cite{Haas98}). A similar structure was found in the 
IRAS maps 
(Habing et al. \cite{Hab84}, \cite{Wal87}, Xu \& Helou \cite{Xu96}), 
in radio continuum surveys 
(Berkhuijsen et al. \cite{Ber83}, Beck et al. \cite{Beck+98}),
in the distribution of OB associations (van den Bergh \cite{Ber91}),
H\,II regions (Baade \& Arp \cite{Baa64}, \cite{Pel78}), and H\,I gas 
(\cite{Sof81}, 
\cite{Cram80}, Brinks \& Shane \cite{Bri84}),
and in several other observations (see \cite{Hod90}). It seems that the
star formation in M\,31 is concentrated on these rings and is very low in the 
inter-ring regions (e.g. \cite{Dev94}, Xu \& Helou \cite{Xu96}). 
Most recently, this has also been suggested from 
mid-infrared spectra observed with ISOCAM (\cite{Ces98}), as well
as from more detailed CO (\cite{Loi+96}, \cite{Loi+98}, \cite{Loi+99}, 
Neininger et al. \cite{Nei+98}) 
and IR (Pagani et al. \cite{Pag+99}) observations of some inner regions of M\,31,
mainly concentrated on the SW half of the galaxy.
Based on the 175\,$\mu$m map the cold dust content has been revealed for the 
total galaxy (Haas et al. \cite{Haas98}). It turned out to be much colder and about 
ten times higher than previously detected. Thus the question arises 
whether we see new, optically highly extinguished regions. 
Especially the ring exhibits numerous knots bright in the FIR.
Their 
luminosity requires an explanation either via the interstellar radiation field 
(ISRF) or via deeply embedded young stars possibly hidden to observations 
at shorter wavelengths.

Therefore, in this paper we concentrate on these discrete FIR sources.
We catalogue the 175\,$\mu$m knots, compare them with 60 and 100\,$\mu$m IRAS data, 
derive their properties, such as colour, luminosity and mass, 
and look for coincidence with radio and optical tracers. 
Finally, the question of star formation in the knots is addressed 
via the $L/M$ ratio, i.e. the normalized energy content,  
and compared to that of the total galaxy as well as the Milky Way and regions 
therein. 

\begin{figure}[tb]
\hspace*{0.45cm}
\rotatebox{270}{\resizebox{2.80cm}{7.95cm}{\includegraphics{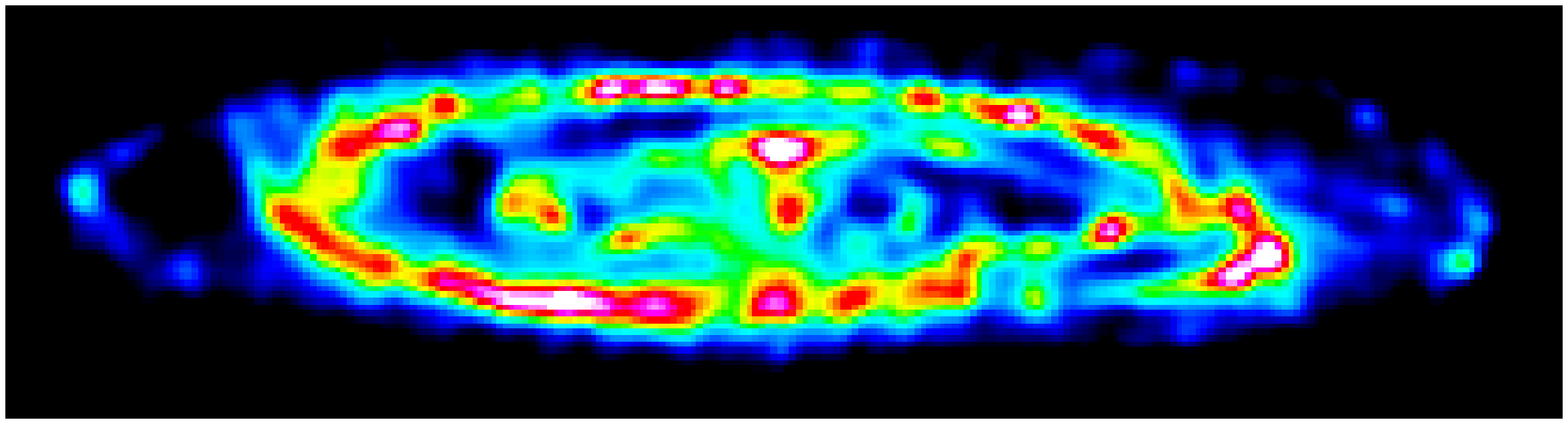}}}
\\[0.5cm]
\rotatebox{270}{\resizebox{3.05cm}{8.7cm}{\includegraphics{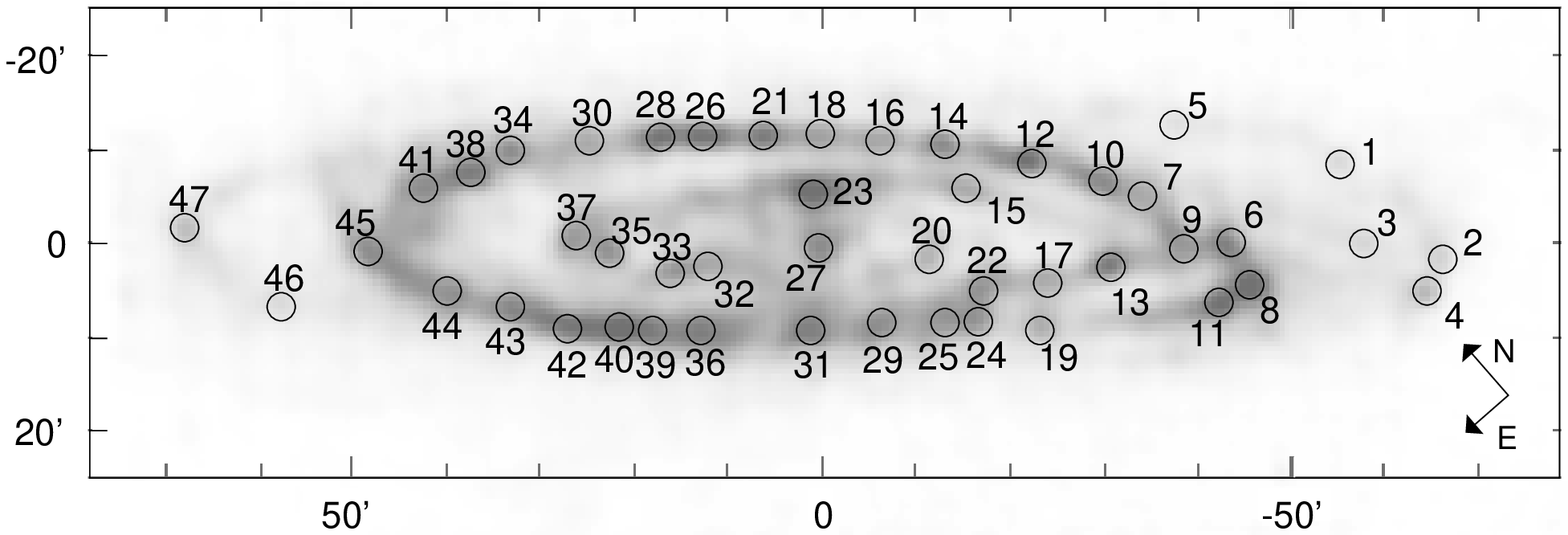}}}
\caption{\label{finding} 
{\it Top:} $175\mu m$ map displaying the ring structure with the
individual knots. 
{\it Bottom:} Finding map displaying the positions of the 47 identified knots.}
\end{figure}
\begin{figure}[t]
\resizebox{8.7cm}{!}{\includegraphics{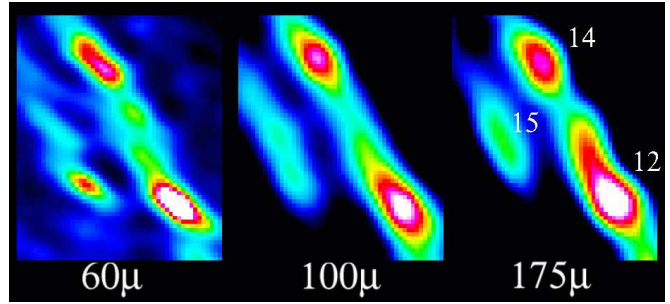}}
\\a) The region around the knots 12, 14, and 15.
\\[0.5cm]
\resizebox{8.7cm}{!}{\includegraphics{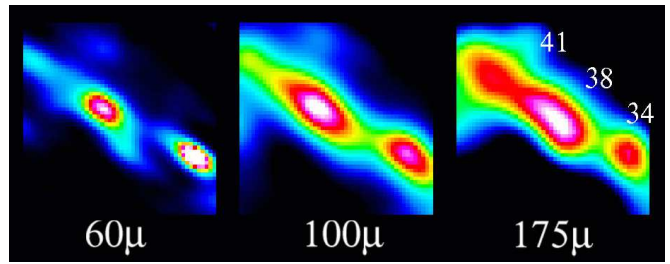}}
\\b) The region around the knots 34, 38, and 41.
\\[0.5cm]
\resizebox{8.7cm}{!}{\includegraphics{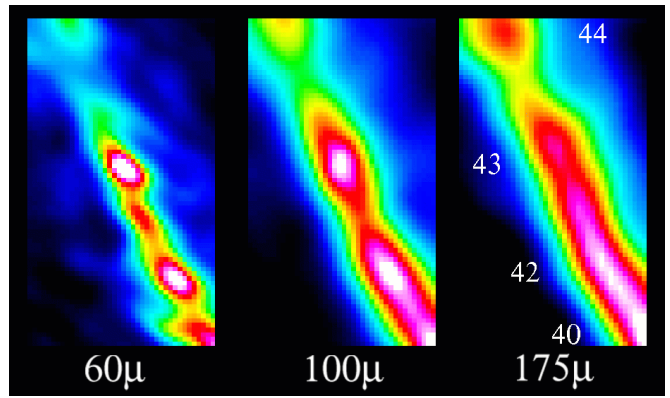}}
\\c) The region around the knots 40, 42, 43, and 44.
\caption{\label{detail} Detailed maps of the areas around some of the knots
at different wavelengths. 
The 60 and 100\,$\mu$m maps are from the IRAS HIRES archive, the 175\,$\mu$m map
s from Haas et al. (\cite{Haas98}). 
}
\end{figure}

\section{Identification and flux determination}
The 175\,$\mu$m map was obtained in February 1997 with ISO's (\cite{Kes96}) 
photometer
ISOPHOT (\cite{Lem96}) covering a field of $3^\circ \times 1^\circ$ 
oriented along the 
major axis at PA $39^\circ$ with a spatial resolution 
of about $1.^\prime 5$. The data reduction and
calibration is described in Haas et al. (\cite{Haas98}).

The discrete sources of the 175\,$\mu$m map 
have first been identified visually,
then two-dimensional elliptical Gaussians have been fitted
over areas of $8^\prime \times 8^\prime$. In total, 47 sources have been 
detected above a threshold of 4 sigma of the local background; they are 
listed in Table \ref{iras}. Columns (2) and (3) give the 175\,$\mu$m
position of the sources with an accuracy of about $15^{\prime\prime}$.
Column (4) gives an approximation of the knot's size derived from the
average of the FWHMs along the two axes. 
The sources are catalogued in right ascension order. 
Fig. \ref{finding} shows the distribution of the knots in a map 
of M\,31, and gives a corresponding finding 
map with the positions for all knots indicated.

\begin{table*}[t]
\caption{The catalogued 175\,$\mu m$ knots with their reference numbers in 
column (1). The  
positions are given in column (2) and (3), an approximated diameter in
column (4). The flux densities at 175\,$\mu m$
have been derived from the ISO map, the corresponding ones at 60\,$\mu m$ and 
100\,$\mu m$ from the high resolution IRAS maps.
Column (8) indicates the classification by
spectral appearance. The temperature of the
cold component, the mass ratio, the total mass, the luminosity ratio and the 
total FIR luminosity have been derived by fitting modified Planck functions 
to the 
spectra. Column (14) gives the number of the corresponding entry 
in the 60\,$\mu$m point source catalogue of Xu \& Helou (\cite{Xu96})
(with uncertain identifications set in brackets), while column (15) 
lists the detections at other wavelengths (see references below).}
\label{iras}
\begin{center}
\scriptsize
\begin{tabular}{ c c c c c c c @{\hspace{0.25cm}}c@{\hspace{0.25cm}} c c@{\hspace{0.3cm}} c@{\hspace{0.25cm}} c@{\hspace{0.25cm}} c @{\hspace{0.1cm}}c@{\hspace{0.2cm}} c@{\hspace{0.03cm}} c@{\hspace{0.03cm}} c@{\hspace{0.03cm}} c@{\hspace{0.03cm}} c@{\hspace{0.03cm}} c@{\hspace{0.03cm}} c@{\hspace{0.03cm}} c }
\hline
 \noalign{\smallskip}
(1) & (2) & (3) & (4) & (5) & (6) & (7) & (8) & (9) & (10) & (11) & (12) & (13) & (14) & \multicolumn{8}{c}{(15)}\smallskip\\
No & RA$_{2000}$ & Dec$_{2000}$ & $d$ & $\displaystyle\rm\frac{F_{175}}{Jy}$ & $\displaystyle\rm\frac{F_{60}}{Jy}$ & $\displaystyle\rm\frac{F_{100}}{Jy}$ & Type & $\displaystyle\rm \frac{T_c}{K}$ & $\rm m_c/m_w$ & $\displaystyle\frac{\rm m_{total}}{10^3m_\odot}$ &$\rm L_c/L_w$ & $\displaystyle\frac{\rm L_{total}}{10^6L_\odot}$ & PS60 & \multicolumn{8}{c}{*)} \\
 \noalign{\smallskip}
\hline
 \noalign{\smallskip}
1  & 0 39 13.8 & 40 36 44 & $3.^{\prime}6$ & ~4.91 & ~1.59 & ~4.03 & ~II & 18.5 & ~152.0 & ~2.702 & ~1.49 & ~2.6 & 6 &        & & & &E&F& & \\
2  & 0 39 22.4 & 40 22 15 & $6.^{\prime}0$ & ~4.83 & ~1.11 & ~3.45 & ~II & 18.2 & ~227.8 & ~2.909 & ~2.02 & ~2.2 & 3 &        & & & &E&F& & \\
3  & 0 39 41.7 & 40 29 17 & $3.^{\prime}6$ & ~3.64 & ~0.41 & ~2.30 & ~II & 18.0 & ~596.5 & ~2.406 & ~4.95 & ~1.4 & 4 &        & & & & &F& & \\
4  & 0 39 43.4 & 40 21 02 & $3.^{\prime}2$ & 11.68 & ~3.31 & 11.49 & III & 21.0 & ~157.1 & ~3.910 & ~3.29 & ~6.2 & 1 &        & &C& &E&F&G& \\
5  & 0 39 51.3 & 40 53 03 & $4.^{\prime}3$ & ~3.73 & ~0.50 & ~2.21 & ~II & 18.0 & ~500.0 & ~2.477 & ~4.15 & ~1.5 & &          & & & & &F& & \\
6  & 0 40 31.4 & 40 41 06 & $4.^{\prime}2$ & 18.68 & ~0.88 & ~8.40 & ~~I & 17.0 & 2300.0 & 16.253 & 13.55 & ~6.0 & (7) &      & & &D& &F&G&H \\
7  & 0 40 37.6 & 40 52 51 & $3.^{\prime}6$ & ~8.95 & ~0.97 & ~5.57 & ~II & 18.0 & ~653.8 & ~6.013 & ~5.43 & ~3.4 & &          & & &D& &F&G& \\
8  & 0 40 39.4 & 40 36 21 & $4.^{\prime}1$ & 28.82 & ~4.16 & 15.64 & ~II & 17.0 & ~538.5 & 24.768 & ~3.17 & 11.1 & 5 &        & & & & &F&G&H \\
9  & 0 40 43.1 & 40 45 51 & $3.^{\prime}7$ & 12.63 & ~0.00 & ~0.16 & ~~I & & & & & & &                                        & & &D& &F&G& \\
10 & 0 40 44.2 & 40 55 51 & $3.^{\prime}3$ & 19.20 & ~1.61 & ~9.88 & ~II & 17.0 & 1043.5 & 16.969 & ~6.15 & ~6.7 & &          & &C&D& &F&G&H \\
11 & 0 40 57.8 & 40 37 52 & $4.^{\prime}5$ & 23.60 & ~3.28 & 14.48 & ~II & 18.0 & ~488.9 & 15.571 & ~4.06 & ~9.3 & &          & & &D&E&F&G&H \\
12 & 0 40 59.9 & 41 04 07 & $3.^{\prime}0$ & 27.12 & ~5.64 & 17.17 & ~II & 17.6 & ~287.2 & 19.138 & ~2.08 & 11.9 & 12 &       & & &D&E&F&G& \\
13 & 0 41 18.7 & 40 49 53 & $3.^{\prime}0$ & 25.83 & ~4.57 & 19.12 & ~II & 19.0 & ~316.7 & 13.463 & ~3.64 & 11.4 & &          & & &D& &F& &H \\
14 & 0 41 23.6 & 41 12 23 & $3.^{\prime}6$ & 18.30 & ~3.60 & 13.40 & ~II & 18.5 & ~283.0 & 10.633 & ~2.77 & ~8.2 & 18 &       A& & &D&E&F&G& \\
15 & 0 41 34.3 & 41 08 08 & $4.^{\prime}7$ & ~9.14 & ~0.44 & ~4.27 & ~~I & 17.0 & 2291.7 & ~7.773 & 13.50 & ~2.9 & &          & &C&D& & & & H\\
16 & 0 41 44.8 & 41 17 54 & $4.^{\prime}9$ & 10.22 & ~2.12 & ~6.96 & ~II & 18.0 & ~290.9 & ~6.804 & ~2.42 & ~4.6 & 20 &       A& & & &E&F& & \\
17 & 0 41 50.4 & 40 54 09 & $6.^{\prime}1$ & 10.71 & ~0.71 & ~4.32 & ~~I & 16.3 & 1454.5 & 11.309 & ~6.66 & ~3.5 & &          & & &D& &F& & H\\
18 & 0 42 03.4 & 41 23 24 & $4.^{\prime}0$ & 11.55 & ~0.72 & ~5.23 & ~~I & 17.0 & 1555.6 & ~9.895 & ~9.16 & ~3.8 & &          A& & &D& & & &H \\
19 & 0 42 10.2 & 40 52 09 & $4.^{\prime}9$ & 10.57 & ~1.86 & ~6.86 & ~II & 18.0 & ~357.1 & ~7.083 & ~2.97 & ~4.6 & 10 &       & & &D& &F& & H\\
20 & 0 42 18.1 & 41 06 24 & $3.^{\prime}5$ & ~8.20 & ~0.91 & ~6.20 & III & 19.5 & ~733.3 & ~3.890 & ~9.84 & ~3.3 & 15 &       & & &D&E& & & \\
21 & 0 42 22.1 & 41 28 09 & $3.^{\prime}5$ & 23.17 & ~4.42 & 15.41 & ~II & 18.0 & ~295.8 & 14.883 & ~2.46 & 10.1 & 25 &       A&B& &D&E&F&G&H \\
22 & 0 42 20.8 & 40 59 09 & $5.^{\prime}0$ & 11.91 & ~0.50 & ~5.97 & ~II & 17.5 & 3250.0 & ~9.185 & 22.79 & ~3.9 & &          & & &D& &F& &H \\
23 & 0 42 30.1 & 41 20 09 & $4.^{\prime}5$ & 31.45 & ~7.36 & 18.48 & ~~I & 16.5 & ~338.5 & 31.171 & ~1.67 & 14.2 & &          & & &D& & & & \\
24 & 0 42 31.4 & 40 58 39 & $5.^{\prime}0$ & 14.81 & ~1.18 & ~5.50 & ~~I & 15.5 & 1611.1 & 20.497 & ~5.45 & ~4.8 & &          & & &D& &F&G&H \\
25 & 0 42 40.7 & 41 00 39 & $4.^{\prime}4$ & 12.40 & ~1.12 & ~5.48 & ~~I & 16.3 & 1117.6 & 13.453 & ~5.12 & ~4.3 & &          & & &D&E&F&G&H \\
26 & 0 42 43.4 & 41 33 24 & $4.^{\prime}5$ & 27.47 & ~5.56 & 20.93 & ~II & 19.0 & ~270.3 & 14.179 & ~3.10 & 12.5 & 26 &       & & &D&E&F&G&H \\
27 & 0 42 52.7 & 41 16 09 & $4.^{\prime}2$ & 17.42 & 19.34 & 30.62 & III & 25.0 & ~~14.4 & ~2.719 & ~0.85 & 19.1 & &          &B&C&D& & & & \\
28 & 0 43 00.8 & 41 37 09 & $3.^{\prime}5$ & 27.12 & ~9.74 & 26.36 & III & 20.0 & ~110.7 & 11.047 & ~1.73 & 15.7 & &          & & &D& &F&G& \\
29 & 0 43 09.9 & 41 05 54 & $3.^{\prime}7$ & 17.75 & ~2.12 & 12.33 & ~II & 19.0 & ~619.0 & ~9.197 & ~7.11 & ~7.0 & 13&        & & &D&E&F& &H \\
30 & 0 43 30.4 & 41 43 08 & $6.^{\prime}2$ & 10.75 & ~1.58 & ~7.01 & ~II & 18.0 & ~434.8 & ~7.080 & ~3.61 & ~4.4 & 30 &       & & & & &F& &H \\
31 & 0 43 35.2 & 41 11 53 & $3.^{\prime}2$ & 22.95 & ~3.92 & 16.08 & ~II & 18.5 & ~345.5 & 13.459 & ~3.38 & 10.0 & (16/17/19)&& & &D&E&F& &H \\
32 & 0 43 39.4 & 41 24 38 & $4.^{\prime}1$ & ~8.04 & ~1.05 & ~7.51 & III & 21.3 & 3700.0 & ~2.614 & 84.3  & ~3.5 & &          & &C&D& & &G& \\
33 & 0 43 55.5 & 41 26 52 & $3.^{\prime}5$ & 13.06 & ~2.10 & ~8.63 & ~II & 18.2 & ~383.3 & ~8.144 & ~3.40 & ~5.4 & 24 &       & &C&D& &F& &H \\
34 & 0 44 01.3 & 41 49 07 & $3.^{\prime}6$ & 15.38 & ~6.80 & 15.97 & III & 20.5 & ~~76.8 & ~5.438 & ~1.39 & ~9.7 & 31 &       & & &D&E&F& & \\
35 & 0 44 10.3 & 41 33 51 & $3.^{\prime}2$ & 13.58 & ~2.67 & ~9.18 & ~II & 18.3 & ~287.5 & ~8.151 & ~2.64 & ~6.0 & 27&        & & &D& &F&G&H \\
36 & 0 44 15.3 & 41 20 36 & $4.^{\prime}5$ & 21.47 & ~4.35 & 11.79 & ~~I & 16.5 & ~407.9 & 21.951 & ~2.01 & ~9.4 & 22&        A& & &D&E&F&G&H \\
37 & 0 44 15.8 & 41 37 36 & $5.^{\prime}1$ & 14.21 & ~0.90 & ~6.29 & ~~I & 16.5 & 1708.3 & 14.489 & ~8.41 & ~4.6 & 29 &       & & &D& & &G&H \\
38 & 0 44 24.2 & 41 51 20 & $4.^{\prime}1$ & 24.06 & ~7.25 & 20.27 & III & 19.0 & ~154.5 & 12.086 & ~1.77 & 12.5 & (33) &     & & &D&E&F& & \\
39 & 0 44 32.8 & 41 25 05 & $4.^{\prime}0$ & 20.40 & ~3.54 & 16.92 & III & 20.0 & ~342.9 & ~8.501 & ~5.36 & ~9.1 & 23&        & & &D&E& &G&H \\
40 & 0 44 40.8 & 41 27 49 & $3.^{\prime}6$ & 30.44 & ~4.31 & 18.92 & ~II & 17.8 & ~491.8 & 21.234 & ~3.82 & 12.1 & &          & & &D&E&F&G&H \\
41 & 0 44 47.1 & 41 54 04 & $6.^{\prime}9$ & 12.65 & ~1.53 & ~5.28 & ~~I & 15.7 & ~851.9 & 16.265 & ~3.11 & ~4.5 & (35)&      & & &D& &F&G& \\
42 & 0 44 58.4 & 41 32 03 & $3.^{\prime}6$ & 23.59 & ~4.69 & 18.49 & ~II & 19.3 & ~266.7 & 11.344 & ~3.36 & 10.7 & &          & & &D& &F&G&H \\
43 & 0 45 10.6 & 41 38 02 & $4.^{\prime}0$ & 18.32 & ~5.01 & 17.54 & III & 20.5 & ~158.3 & ~6.753 & ~2.87 & ~9.5 & 28 &       & &C&D& &F&G&H \\
44 & 0 45 28.3 & 41 44 45 & $5.^{\prime}2$ & 16.38 & ~2.06 & 10.24 & ~II & 18.0 & ~574.1 & 10.967 & ~4.77 & ~6.4 & &          & & &D& &F&G&H \\
45 & 0 45 35.5 & 41 54 14 & $5.^{\prime}4$ & 15.69 & ~1.07 & ~8.29 & ~II & 17.5 & 1416.7 & 12.016 & ~9.93 & ~5.4 & &          & & &D&E&F&G&H \\
46 & 0 46 32.2 & 41 59 08 & $3.^{\prime}7$ & ~5.23 & ~0.59 & ~2.70 & ~II & 16.7 & ~760.9 & ~4.951 & ~4.03 & ~1.9 & 36 &       & & & & &F& &H \\
47 & 0 46 33.0 & 42 11 37 & $3.^{\prime}0$ & 11.01 & ~2.95 & ~8.00 & ~II & 18.0 & ~208.3 & ~7.097 & ~1.73 & ~5.4 & 39 &       & & & &E&F& & \\
\noalign{\smallskip}
\hline
\end{tabular}
\end{center}
*)References for Column(15):\\
(A) 5C3 survey of sources at 1421\,MHz (\cite{Gil79});
(B) 6\,cm HEASARC-NORTH catalogue; 
(C) Supernova remnants at 6\,cm (\cite{Dic84});
(D) Dark clouds listed in Hodge (\cite{Hod80});
(E) Radio sources from the 36W catalogue at 610\,MHz, 
thought to be H\,II regions (Bystedt et al. \cite{Bys84});
(F) H\,II regions from an H$\alpha$ survey of M\,31 (\cite{Pel78})
(G) CO measurements from Dame et al. (\cite{Dam+93}); 
(H) H\,I measurements from Brinks \& Shane (\cite{Bri84}).
\end{table*}

\begin{figure*}
\centerline{
\rotatebox{270}{\resizebox{4.0cm}{5.7cm}{\includegraphics{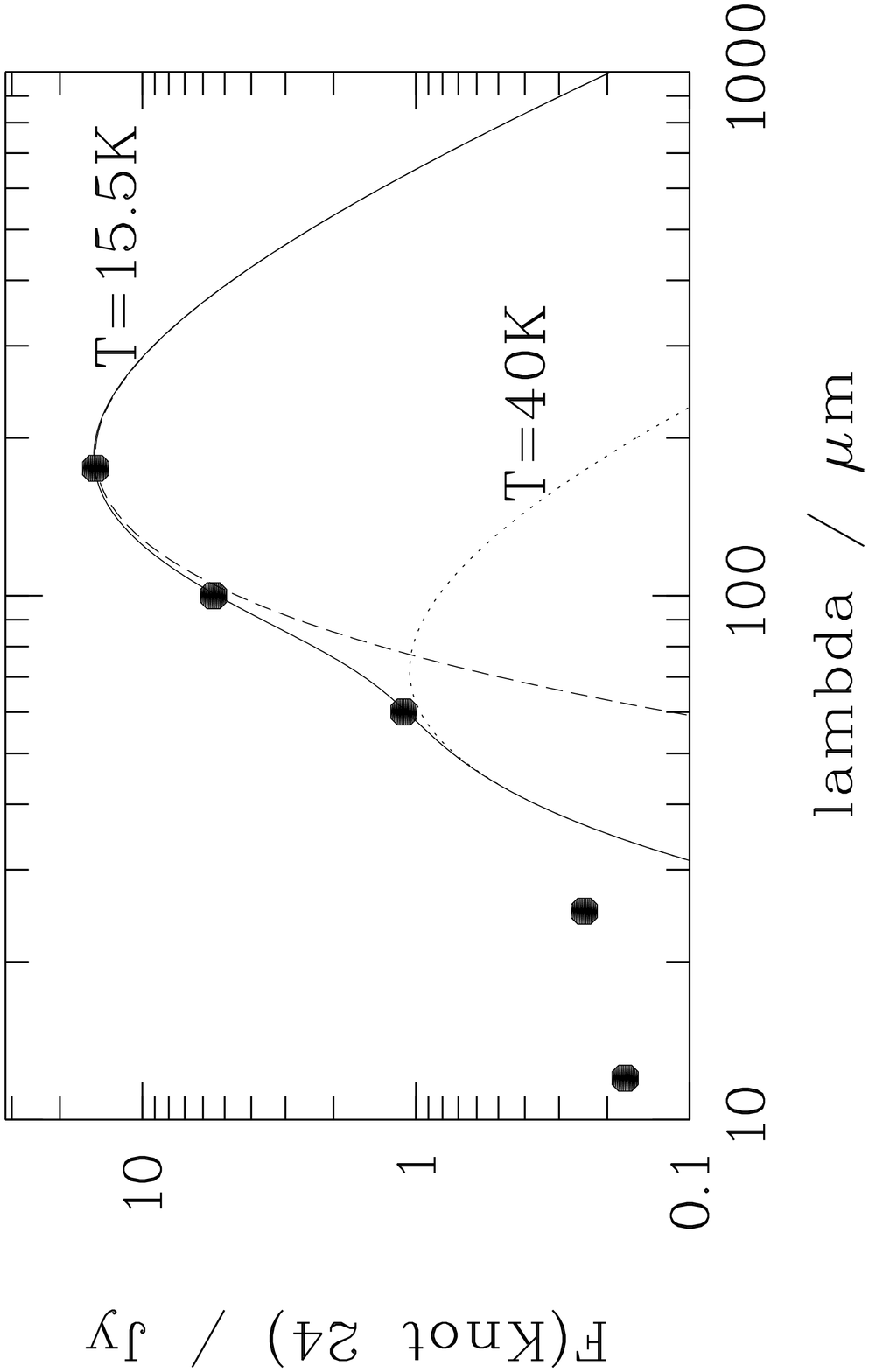}}}
\hspace*{0.25cm}
\rotatebox{270}{\resizebox{4.0cm}{5.7cm}{\includegraphics{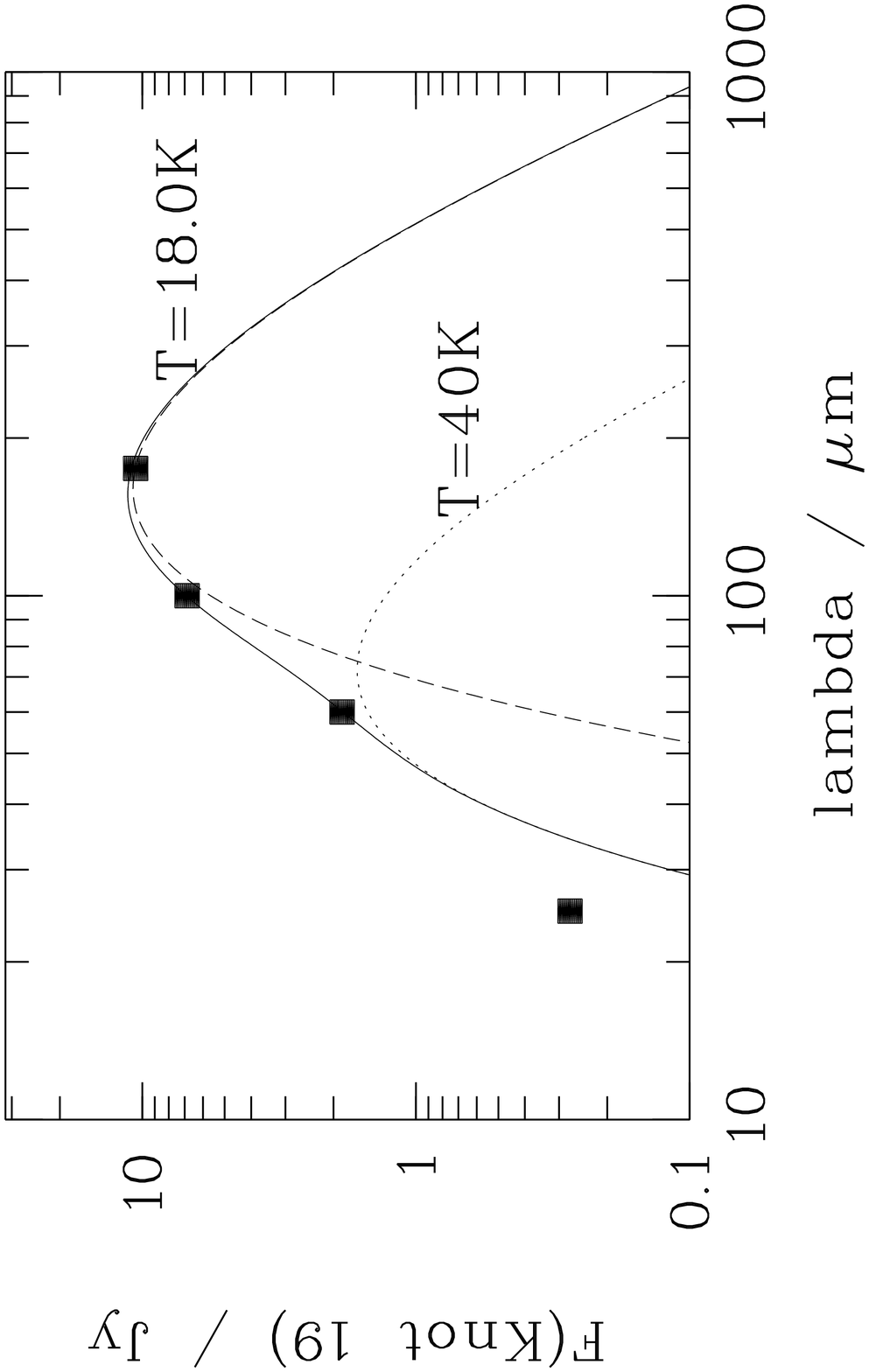}}}
\hspace*{0.25cm}
\rotatebox{270}{\resizebox{4.0cm}{5.7cm}{\includegraphics{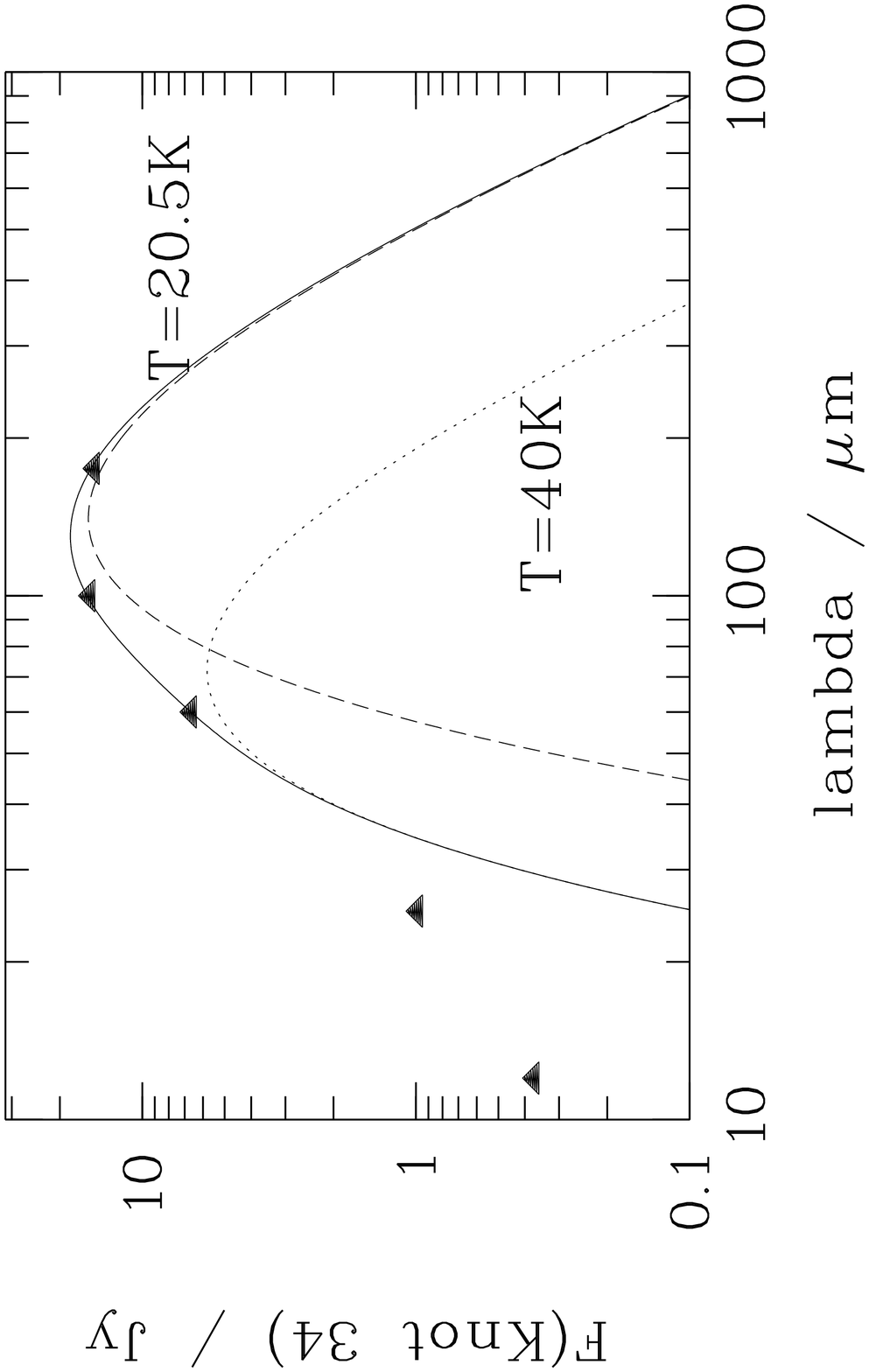}}}}
\caption{\label{spectra} SEDs of three typical knots: type 
I = cold (left), II = medium (middle), III = warm (right). 
The flux errors are smaller than the symbol sizes. The lines 
indicate modified Planck curves (emissivity $\propto \lambda ^{-2}$).}
\end{figure*}

Comparing these knots with the list of 60\,$\mu$m point sources in M\,31
(39 entries, Xu \& Helou \cite{Xu96}) yields no complete equivalence
(see column (14) of Table \ref{iras}). In general, more
discrete sources are found at 175\,$\mu$m, but there are also point
sources at 60\,$\mu$m, which are not
prominent at 175\,$\mu$m. Except for Xu \& Helou's point source \# 2, which is
a background galaxy 
located outside the 175\,$\mu$m map, all of these point sources are
faint at 60\,$\mu$m. However, at their position we do find
some brightness enhancement in the 175\,$\mu$m map but no source above
the 4\,$\sigma$ range is resolvable. The same holds for the 175\,$\mu$m 
sources, that are not included in the list of Xu \& Helou.
They do all show up in the 60\,$\mu$m map, but are either too faint or too 
extended 
to have been included in Xu \& Helou's list of point sources.

A close inspection of the areas around the knots shows that 
most of the 175\,$\mu$m knots coincide well with their 60\,$\mu$m counterparts 
(e.g. Fig. 2a, Knot 12, Knot 14), but small 
shifts between the 60\,$\mu$m and 175\,$\mu$m peaks are also observed
(e.g. Fig. 2a, Knot 15, Fig. 2b, Knot 31, Fig. 2c, Knot 43). Fig. 2c also 
shows an example of a distinct source at 60\,$\mu$m, which is not prominent
at longer wavelengths but located in the
dip between two 175\,$\mu$m knots (42 and 43), whereas Fig. 2b 
shows Knot 41 at 175\,$\mu$m,
which is too faint to be recognized at 60\,$\mu$m. 
These shifts and also the additional sources prominent at only one wavelength
 can be explained by a smooth cold 
dust distribution which is just heated at several points yielding the
60\,$\mu$m peaks, and with the 175\,$\mu$m sources as the remaining less heated
parts. A situation like this would be expected in a complex of dust clouds 
and H\,II regions, where only the dust close to the ionized regions reaches
high temperatures. 
However, these shifts between 60\,$\mu$m and 175\,$\mu$m peaks are rare and 
do not affect the basic conclusions drawn from the following investigations.

The total flux densities of the knots have been determined by 
aperture photometry with standard routines of MIDAS and DAOPHOT, whereby 
the local background within M\,31 has been subtracted. The flux 
density is sensitive to the size of the chosen aperture as the sources 
are not completely separated but situated very close together and on top of 
the ringlike structure. 
It has been measured within an aperture of $5^\prime$, the background
in the surrounding $0.5^\prime$ wide ring. 
The aperture has been chosen as being large enough to measure a 
sufficient area of the knots without being too much affected by 
the varying background.

The flux densities of the sources are
given in column (5) of Table \ref{iras}.
Their values vary between 3 and 32\,Jy.
Note that at 175\,$\mu$m, the  brightest source
(Knot \# 23) is not at the position of the galaxy's centre. Instead it is
situated about $5^\prime$
to the NW and is part of the innermost of the concentric rings.
It will later be discussed in Sect. \ref{shl23}.

The flux densities at 12, 25, 60 and 100\,$\mu m$
have been determined from the high resolution IRAS maps smeared to the 
spatial resolution of $1.^\prime 5$ as for the ISO 175\,$\mu$m data. 
Special care has been taken to use the same aperture of $5^\prime$ for 
all passbands to allow reasonable comparison.
The flux values of the longer, more relevant wavelengths are listed as 
columns (6) and (7) in Table \ref{iras}.

\section{Spectral Energy Distributions (SEDs)}
Since the 12 and 25\,$\mu$m 
are dominated by a different kind of dust (small transiently heated grains, 
according to D\'esert et al. (\cite{Des+90}) and Dwek et al. (\cite{Dwek+97})), 
which does not show up at 175\,$\mu$m, 
we concentrate on the 60 -- 175\,$\mu$m SEDs.
They show a continuous range of shapes. 
For the purpose of recognising trends we have binned the SEDs into  
three different types, an example for each is given 
in Fig. \ref{spectra}. The spectra of type I show a rising flux density 
with increasing wavelength in a steep 
slope without any curvature. This indicates a very low temperature, since the 
maximum of any Planck curve is barely reached at 175\,$\mu$m. 
Type II is the most common case, still with a monotonous rising flux density,
but with a clear curvature. Type III
contains those spectra where the flux density reaches a 
maximum value within the observed spectral range. 

The comparison with the temperature of the cold dust component
(Sect. \ref{temp_sec}) shows in fact that the SED types are  
well correlated to this temperature. Hence we get an unbiased definition
of the types by using $T \le 17$\,K ($\Rightarrow$ I),  
17\,K $ \leq T \le 19$\,K ($\Rightarrow$ II), $T \ge 19$\,K 
($\Rightarrow$ III).
The type of each individual source is indicated in 
column (8) of Table \ref{iras}.

\section{Warm and cold dust component}
\label{temp_sec}

For further investigation of the knots, modified Planck curves have been fitted to the 
SEDs. We assumed an emissivity law of $Q_\lambda \sim \lambda^{-2}$.
Only the points at 60, 100, and 175\,$\mu$m have 
been fitted. 
Except for Knot 32 (close to the centre), all knots require (at least)
two dust components with different temperatures to fit the data. 

In order to fit two modified Planckians four parameters 
(two temperatures and two intensities), 
have to be determined
with three data points now available (at 60, 100 and 175\,$\mu$m).
Therefore the temperature of the warmer component had to be fixed.
A temperature of 40\,K has been chosen in accordance with 
previous investigations (\cite{Wal87}). 
Note, that the value itself has
no physical meaning, as the flux density at 60\,$\mu$m is already influenced
by transiently heated very small dust grains.
However,
when hold constant for the whole knot sample, it allows for a reasonable 
parameterized comparison of the sources and their dust properties.


\section{FIR colours of the knots}

\begin{figure}
\rotatebox{270}{\resizebox{!}{8.7cm}{\includegraphics{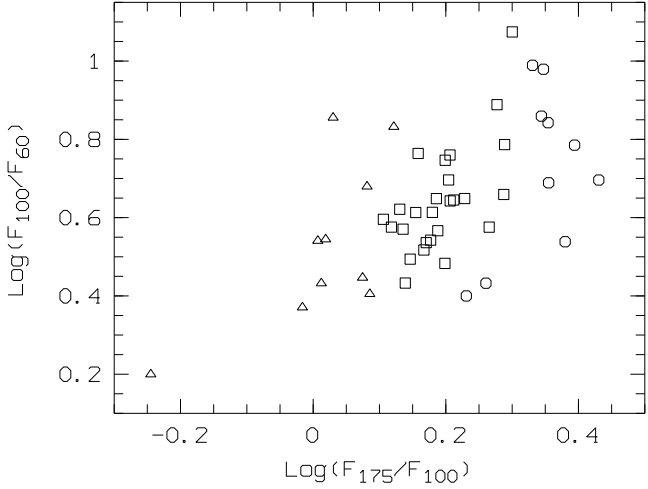}}}
\caption{\label{col_diag} FIR colour -- colour diagram:
The colour $\log(F_{100}/F_{60})$ is plotted against $\log(F_{175}/F_{100})$  for each knot. 
The three SED types are indicated by different symbols: 
$^{_\bigcirc}$ = I (cold), $\Box$ = II (medium), $\triangle$ = III (warm).}
\end{figure}

In order to visualize the spectral shape of all knots at once  
Fig. \ref{col_diag} shows the colour-diagram, where $\log(F_{100}/F_{60})$ is
plotted against $\log(F_{175}/F_{100})$. For each SED type a different 
symbol has been chosen for better identification.
The types are clearly distributed on different regions in the diagram,
although they do not separate completely.
The colour-diagram emphasizes the gain, provided by the 175\,$\mu$m data.
Whereas the knots are not distinguished along the $\log(F_{100}/F_{60})$ axis,
the inclusion of 175\,$\mu$m yields the separation of the types. 
$\log(F_{175}/F_{100})$ is actually  a colour temperature indicator. 

The triangle in the lower left corner of Fig. \ref{col_diag} belongs to 
Knot 27, which is the nucleus of M\,31 having the bluest colour and therefore being 
the warmest knot, as already discussed by Habing et al. (\cite{Hab84}) and 
Hoernes et al. (\cite{Hoe+98}). 
From a single modified Planck fit to the SED, its temperature is estimated at about 
29\,K ($\lambda^{-2}$ -emissivity law). 
As the 175\,$\mu$m map gives 
no new information for such warm dust in addition to the 60\,$\mu$m and 100\,$\mu$m, 
we exclude this source from our further discussion. 

\section{Luminosities}
\label{sec_l}

The IR luminosities have been determined by integrating (1) the modified Planck curves
between 8\,$\mu$m and 1000\,$\mu$m, and -- for comparison -- (2) the measured 
fluxes 
in the three rectangular bandpasses.
The latter values are about 10\% lower, as expected due to the missing 
fraction at longer wavelengths: beyond 230\,$\mu m$, the flux density is 
extrapolated by the modified Planck curve. Columns (12) and (13)
of Table \ref{iras} list the luminosity ratio of cold and warm modified Planckian and the total FIR luminosity
within the chosen aperture. 

The FIR luminosity integrated over all knots yields 
$2.98\cdot 10^{8}L_\odot$, which is about 12\% of the FIR 
luminosity estimated by Haas et al. (\cite{Haas98}) for the whole galaxy. 
However, this is only a lower limit, as the apertures cut off some
of the flux of each knot.
An upper limit can be estimated as follows: 
the rings comprise about 30\% of the total luminosity of the whole galaxy, 
whereby 5\% may be attributed to the disk. 
The luminosity of the knots
is therefore less than 25\% of that of the whole galaxy.

\section{FIR colour luminosity diagram}

\begin{figure}[tb]
\rotatebox{270}{\resizebox{!}{8.7cm}{\includegraphics{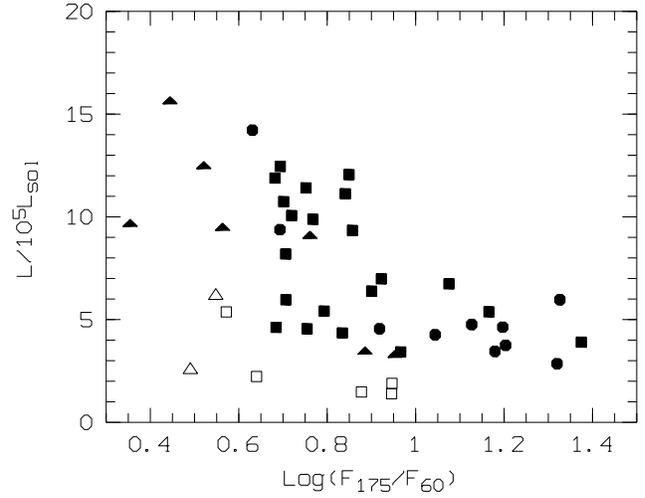}}}
\caption{\label{hr} FIR luminosity against $\log(F_{175}/F_{60})$  colour. 
Filled symbols belong to knots in the ring and 
inside, open ones to the knots in the outer parts of the galaxy.}
\end{figure}

Fig. \ref{hr} shows the FIR luminosity plotted against the colour 
$\log(F_{175}/F_{60})$. 
The most luminous knots are also blue and warm, while the cold knots 
seem to be less luminous.  
Whereas the points are more or less evenly distributed in the left-hand and 
lower half of the diagram, none are found for red colours and high 
luminosities. A statistical bias in our knot sample, due to the 
apertures cutting extended flux, could produce this empty 
region to the upper left of the figure
if red knots are generally further extended than blue knots.
A check on the sizes of the knots revealed, however, that the extension of the
knots does not correlate with the colour.
Therefore we rather think that the trend in Fig. \ref{hr} is real. 
An explanation may be that for
a cloud with high luminosity the mass is also high (see 
Eqn. \ref{eqn_m_l} below), making it very probable that star
formation has already started and shifted the colour towards the blue.

The knots situated outside the 10\,kpc ring  
have a low luminosity, but show all blue or medium colours, suggesting that even in 
the outermost regions of the galaxy ongoing star formation can be found 
(see Sect. \ref{nature}). 
Interestingly, although the knot sample was 175\,$\mu$m selected, 
we did not find red knots in the outer regions. 

Finally, the knots situated in the bright ring and inside of it seem to follow a sequence. 
This, however, might be partly due to the flux limit of our sample. 
Only those knots which are prominent enough can be identified within 
the relatively bright and crowded ring.  
Then the apparent colour-luminosity trend is expected, 
as for a given 175\,$\mu$m flux those knots with 
bluer colour have to be more luminous. Nevertheless, our knot sample 
does not seem to be biased in favour of or against cold knots.

\section{Dust masses}
\label{sec_m}
For each knot we determined the dust mass M$_{\rm d}$, adopting the 
model parameters of Chini et al. (\cite{Chi86}): 
\begin{equation}
\label{eqn_m_1}
M_d = \frac{1}{\kappa_0} \frac{D^2 S_{\nu}}{B_{\nu}(T)}\\
\end{equation}
where D is the distance, $S_{\nu}$ the flux density, 
$B_{\nu}$ the Planck function, and
$\kappa_0$ the mass absorption coefficient of the dust at a 
reference wavelength. The fit of the modified Planck functions to the data 
yields equations
\begin{equation}
\label{eqn_snu}
S_\nu = (\frac{125\mu}{\lambda})^2 (I_c B_\nu(\lambda,{\rm 40K}) 
                                       + I_w B_\nu(\lambda, T_w))\\
\end{equation}
where $I_{c}$ and $I_{w}$ are the scaling factors of the cold and warm 
component for each knot. Assuming the same value of $\kappa_0$ for the two
components, this yields
\begin{equation}
\label{eqn_m_d}
M_{d}  = \frac{1}{\kappa_0} D^2 (\frac{125\mu}{\lambda})^2 (I_{c}+I_{w})\\
\end{equation}

We adopted $\kappa$$_{\rm 0}$\,=\,0.03\,m$^{\rm 2}$/kg for a reference
wavelength of 1.3\,mm, which has already been successfully used
for star forming regions and cold cloud fragments (\cite{Chi87}, \cite{Kru94}).
This value is well in between those derived from other 
investigations and computed to 1.3\,mm, 
i.e. $\kappa_{0}\,=\,0.025\,\rm m^2/kg$ from Sodroski et al. (\cite{Sod+94})
(for DIRBE/COBE FIR data) and
$\kappa_{0}\,=\,0.036,\rm m^2/kg$ from Fich \& Hodge (\cite{Fich91}) 
(for IRAS and mm data). 
For the distance of M\,31 we took $D=690\,\rm kpc$. The resulting 
mass values are given in column (11) of Table \ref{iras}. 

Note, that these 
values are only lower limits of the dust masses, as dust with temperatures 
below about 12\,K cannot be detected at 180\,$\mu m$ but only at longer 
wavelengths. The presence of such cold dust has 
already been found in Galactic molecular cloud complexes
(\cite{Kru94}, \cite{Ris+98}) as well as in star forming regions 
(\cite{Ris+99}), and might therefore also be common in M\,31's FIR sources.
However, 
for the Chameleon region
Toth et al. (\cite{Toth+00}) have shown that the number 
of such cold protostellar
cores is very low, and their mass does not contribute significantly to the
total mass of the molecular cloud complex. 
Hence, we assume that in case there is any dust below 12\,K in M\,31
the increase of the derived masses
will not be very high nor critically influence our discussions.

The integrated mass of all knots (each seen within  
5$\arcmin$ aperture) is $5.5\cdot 10^5M_\odot$, which is only about 
6\% of the total dust mass of M\,31 (10$^7M_\odot$, when recalculated 
with the same model parameters; Haas et al. (\cite{Haas98}) 
computed the dust mass 
slightly differently according to Hildebrandt (\cite{Hild83})). 
It will be interesting to compare the dust masses of M\,31 with the masses
of the involved gas. For the southwest part of the galaxy, 
Pagani et al. (\cite{Pag+99})
have found an excellent correlation
between the column density of neutral gas and the mid IR intensity.
Apparently, the gas of the 10\,kpc ring is dominated
by H\,I whereas little H\,I is found inside, in the inter-ring region. 
However, the conversion factor between CO and H$_2$ is not very well
known which might change the total column density of neutral gas. 
Thus, the comparison of M\,31's dust and gas masses derived from
H\,I and CO surveys will be performed in a future detailed work.

\section{The L/M ratio as parameter to measure the normalized energy content}
\label{sec_l_m}

\begin{figure}
\rotatebox{270}{\resizebox{!}{8.7cm}{\includegraphics{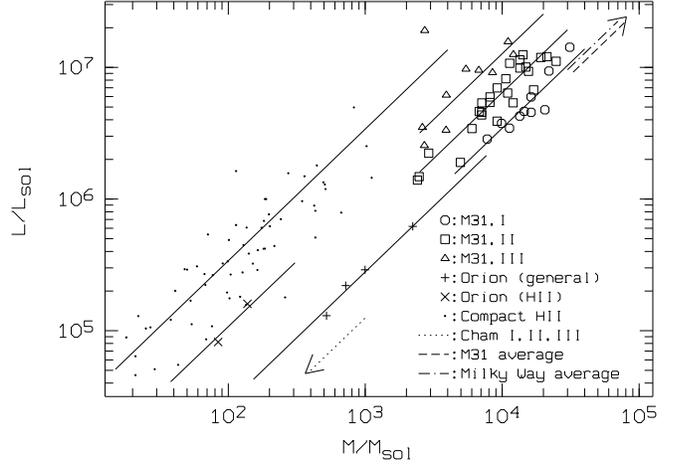}}}
\caption{\label{m_l} FIR luminosity against 
dust mass. The symbols refer to the three 
knot types and to comparison regions as indicated. 
For each group, the linear fit is shown by the lines, with the offsets 
illustrating the different $L/M$ values.
For the Chameleon clouds and the whole galaxies the data values  
lie outside the box, thus the positions of their average lines 
are indicated by arrows. The values and references
are given in Table \ref{l_m_tab}.
}
\end{figure}

\begin{table}
\caption{The FIR luminosity to dust mass ratio 
in solar units averaged for the knots and comparison objects. 
The data for Orion have been recalculated with our model parameters.}
\label{l_m_tab}
\begin{tabular}{l c}
\hline
 \noalign{\smallskip}
Objects & $L_{\rm FIR}/M_{\rm Dust}$ \\
 \noalign{\smallskip}
\hline
M\,31 knots SED type I               &  345 $\pm$  76 \\
M\,31 knots SED type II              &  642 $\pm$ 159 \\
M\,31 knots SED type III             & 1270 $\pm$ 311 \\ 
M\,31 (average)                      &  280 $\pm$  56 \\
M\,31 (10\,kpc ring without knots)   &  267 $\pm$  35 \\
M\,31 (5\,kpc ring without knots)    &  271 $\pm$  47 \\
M\,31 (disk, 8--12\,kpc distance)    &   26 $\pm$   4 \\
M\,31 (disk, 3--6\,kpc distance)     &   45 $\pm$   7 \\
M\,31 (central part 0.8--1.3\,kpc)   &  145 $\pm$  34 \\
Orion (general) (\cite{Wall96})      &  284 $\pm$  59 \\
Orion (H\,II)  (\cite{Wall96})       & 1070 $\pm$ 217 \\
Compact H\,II regions (\cite{Chi87}) & 3396 $\pm$ 264 \\
Chameleon (Boulanger et al. \cite{Boul98})            &  151 $\pm$  25 \\
Milky Way (average) (Sodroski et al. \cite{Sod+94})   &  320 $\pm$  32 \\
 \noalign{\smallskip}
\hline
\end{tabular}
\end{table}

In order to compare knots in M\,31 and dust clouds 
in the Milky Way, a suitable parameterisation is required. 
We use the $L/M$ ratio as parameter. The advantage over 
using the temperature alone lies in the fact that $L/M$ accounts simultaneously for 
an increase of $L$ due to the warm dust and an increase of $M$ due to the 
cold dust. Note that the 175\,$\mu$m data now allow for a reasonable 
estimate of the cold dust 
component, which was not possible from the 60 and 100\,$\mu$m data alone. 
In the case of one single modified Planckian, $L/M$ is equivalent to using the temperature 
(see Eqn. \ref{eqn_m_1}). So far the $L/M$ ratio provides 
a measure for the normalized energy content of the dust. 
Power sources are the interstellar radiation field (ISRF) 
and star formation (SF) in the knots.
Note that our $L/M$ ratio considers only the dust, thus it differs from 
previous concepts
of L$_{\rm FIR}$/M$_{\rm GAS}$ derived for galaxies, compact H\,II regions, 
and various molecular clouds and star forming regions 
(Chini et al. \cite{Chi86}, \cite{Chi87}, \cite{Wall96}, Boulanger et al. \cite{Boul98}).
We will first consider only SF. 

In Fig. \ref{m_l} the FIR luminosity of each knot is plotted against
its dust mass. The knots fill a continuous range in this diagram, but 
the three SED types are well separated. 
A least square fit
\begin{equation}
\label{eqn_m_l}
L = C \cdot M^{\rm \alpha}, \\
\end{equation} 
yields $\alpha$ = 1.03\,$\pm$\,0.10 but a different value $C$ for each SED-type.
With  $\alpha$ being one, $C$ is equal to the average $L/M$-ratio, listed in 
Table \ref{l_m_tab}. Thus within each SED type the $L/M$ ratio
seems to be independent of the absolute values of $L$ or $M$, which allows for
a direct comparison of large and small cloud complexes via $L/M$.

In order to disentangle the power contribution of the ISRF and SF 
in the knots, we try to find some 
reasonable values
for the ISRF. 
One can safely assume that the regions outside the rings
are mainly heated by the ISRF.  
This is equivalent to the assumption that in M\,31 the SF is confined to 
the three ring-like structures at radii of 5, 10 and 14\,kpc which is
consistent with H$_{\alpha}$ observations (\cite{Dev94}).
As the ISRF variates along the galaxy, i.e. it is higher in the bulge 
region than in the outer parts (Pagani et al. \cite{Pag+99}),
we have determined the $L/M$-ratios for several fields in the following 
regions: 
\begin{itemize}
\item[(1)] in the central bulge region (0.8-1.3\,kpc)
\item[(2)] in the disk between 3 and 6\,kpc, excluding the 5\,kpc ring 
\item[(3)] in the disk between 8 and 12\,kpc, excluding the 10\,kpc ring
\item[(4)] along the 5\,kpc ring but avoiding the knots. 
\item[(5)] along the 10\,kpc ring but avoiding the knots. 
\end{itemize}
The resulting average values for these regions 
are given in Table \ref{l_m_tab}. Except for the central part, the
$L/M$-ratio of the inter-ring regions are extremely low. In the rings,
instead, the average value outside the knots $L/M = 269$ is quite high. 
This might 
be explained by an increase of the ISRF in the ring 
or more likely by the assumption that the ring contains more unresolved dust 
clouds and/or star forming region.

Inside the knots the ISRF cannot be stronger, 
rather, the knots might be shielded against radiation from 
outside. 
Hence, for dust heated only by the ISRF, 
values between 20 and 50 are expected for the $L/M$ ratio.
The power of the warm and probably also the medium knots 
(having $L/M$ ratios of more than 600) is therefore certainly 
dominated by SF. 
The cold knots have an average $L/M$ ratio of 345, which is still too high
for a heating of the ISRF only, but the 
difference to the surrounding inner-ring regions is small. 
This can be interepreted in the way that in these sources also the 
ISRF plays a major role, which must then be higher in the rings than outside.
However, it does not account for the whole power.
Hence, even the cold knots require additional heating,
which could come from a high number of
low mass stars or from young massive stars. 
Their UV radiation, however, has to be absorbed effectively within the clouds, 
since signatures of a strong UV field are not observed (\cite{Ces98} and 
Pagani et al. \cite{Pag+99}).

\section{Comparison with radio and optical tracers}
\label{tracers}

The celestial coordinates of the knots have been used for identification of 
the sources in observations at other 
wavelengths.
Special emphasis has been placed on the
catalogue of optically identified H\,II regions (\cite{Pel78}) and 
the radio continuum as tracer 
for H\,II regions. We have used the 610\,MHz catalogue of Bystedt et al. 
(\cite{Bys84}) which 
is flux limited and provides a list of well identified sources. In order to 
separate the thermal and nonthermal radio emission, maps at 2\,cm or 6\,cm are
required but are not yet available. 
Furthermore, 
catalogues of supernova remnants (SNRs)
(\cite{Dic84}) and dark clouds (Hodge \cite{Hod80}), and surveys of 
CO (Dame et al. \cite{Dam+93}) and H\,I (Brinks \& Shane \cite{Bri84})
have been looked up for comparison. 
We limited our comparison to these catalogues, because 
they cover the {\it complete} area of M\,31, where the FIR knots are situated.
The results of this search are listed in 
column (15) of Table \ref{iras}. 

These tracers do not simply correlate with classification of the FIR knots.
Nevertheless, some statistical trends can be recognized:
A large number of our warm knots (SED types III, and also II) are identified 
as 
H\,II regions. This is expected, as nearly 
all the 60\,$\mu$m point sources coincide with H\,II regions 
(Xu \& Helou \cite{Xu96}). 
The 20\,cm map
of Beck et al. (\cite{Beck+98}) has been looked up and yields coincidence 
with nearly every
knot independant of the SED type. However, the cold knots of type 
I are not as prominent at 20\,cm as the warm knots of type III.

\begin{figure}
\vspace*{-2.3cm}
\resizebox{9.0cm}{!}{\includegraphics{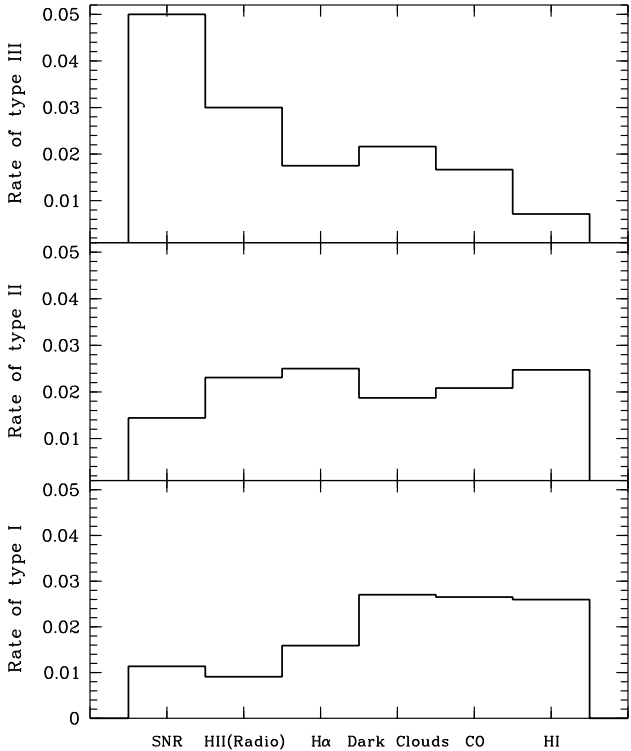}}
\caption{\label{histo} Statistical coincidence of the FIR knots with
detections by other tracers 
separated for each SED type. The cold knots (type I) are 
more related to detections in H\,I and CO, whereas the  warm knots (type III)
are dominant at SNRs and H\,II regions. The medium knots (type II) follow more a mean distribution.
}
\end{figure}

Although the optically dark dust lanes in M\,31 generally match the FIR ring 
quite well, only 
a weak correlation with the dark clouds (Hodge \cite{Hod80}) is  found.
The high number density of these dark 
clouds, however, considerably increases the likelihood of finding one at any randomly 
picked position within 
our spatial resolution of 90$^{\prime\prime}$. In this 
respect, it appears more interesting that a number of even large 
dark clouds like D76, D144, D425 or D517 are not found in the 
175\,$\mu$m map. They probably do not have sufficient mass 
to provide a prominent 175\,$\mu$m emission. 
This picture is consistent with the lack of any correlation between 
optical extinction maps as created by Nedialkov (\cite{Ned98}) and 
our FIR sources.
A moderate amount of just 1 mag of visual foreground extinction already 
results in a dark lane.  

SNRs are found close to the position of some of the knots, but they are 
not sufficiently coincidental to provide evidence for a correspondence. However, 
SNRs are only found for warm (and medium) knots with 
temperatures of the cold dust component above 17\,K (SED types II and III). 
Most of the warm knots correspond with H\,II regions, hence they contain 
high mass stars, i.e. the precursors of SNRs. Therefore 
SNRs are also most likely to be found in warm knots.

In Fig. \ref{histo} the coincidence of the FIR knots with the other tracers is
illustrated as a histogram for the three SED
types.  Whereas the knots of type II more or less follow the general 
distribution, the warm knots are more frequently represented by the SNRs and H\,II 
regions as discussed above, and the cold
knots of type I correlate with detections in CO and H\,I.

For the southwest part of M\,31, new CO observations with higher resolution and
sensitivity have been presented by Neininger et al. (\cite{Nei+98})
and Loinard et al. (\cite{Loi+99}). A 
comparison of the knots of this area with the new CO data reveals, that all of
them have a faint CO counterpart. The same will probably be the case for the 
other radio tracers, where so far no better data are available. The 
catalogues used are intensity limited. Hence, the coincidences 
 of the FIR knots with the other tracers must not be interpreted in the way, 
that e.g. the warm knots do not contain any CO, but that they emit too faintly
to be detected in the survey of Dame et al. (\cite{Dam+93}). On the other hand, also the cold
knots do probably contain H\,II, but the emission is too faint to be detected.
However, this does not alter the general conclusion, that the cold knots are 
more related to CO and H\,I, the warm knots more to H\,II.

\section{Nature of the knots}
\label{nature}

What are the different types of knots? 
In this section we combine the information collected so far: 
the spatial coincidence with other tracers of known physical properties, 
and the $L/M$ ratio tracing the energy content and providing 
unbiased clues for the dominating power source. 
But first we have to consider the spatial extension.

\subsection{Extended cloud complexes}
\label{complexes}
The extension of the knots (listed in column (4) of Table \ref{iras}) 
is significantly higher than the 
FWHM of point sources due to ISO's resolution.
With the adopted distance of 690\,kpc the average diameter computes up
to about 800\,pc. 
Since this appears too large for a single dust cloud, the FIR knots might 
rather represent numerous clouds in chance projection or 
giant complexes of molecular clouds with associated H\,II 
regions, where the single clouds are too small to be separated within
our resolution (300\,pc). 
This consideration is supported by the work of Loinard et al. (\cite{Loi+99}). 
In their
figures 6 e-g, they demonstrate very clearly the influence of angular 
resolution on the observed size of objects, which consist of several small,
unresolved clouds.
The additional possibility of chance projection does not alter our 
conclusions, so in the further discussion we consider only the case of cloud complexes. 

This picture explains well the high dust masses and luminosities which are 
more 
typical for a dwarf galaxy than for a galactic complex. 
For example, the fairly large Orion 
complex in our own Galaxy, including Orion\,A, Orion\,B and $\lambda$\,Ori, 
has a diameter of less than
300 pc, and is therefore a factor of three smaller
than most of the M\,31 knots. Accordingly, for the Orion 
complex the integrated dust mass of $\rm 2.9 \cdot 10^{3} M_\odot$, and
the combined infrared luminosity of $\rm 9.4 \cdot 10^{5} L_\odot$ 
(\cite{Wall96}) are smaller than for the M\,31 knots. 

\subsection{Warm knots: Huge star forming complexes}

The investigations provide evidence that the warm knots of SED type III 
represent huge complexes mainly powered by high mass star formation. 
The reasons are (1)
the frequent spatial coincidence with H\,II regions (Fig. \ref{histo}) and 
(2) the $L/M$ excess well above the ISRF traced by the rest of M31 
(Fig. \ref{m_l}, Table \ref{l_m_tab}). 

This excess is similar to that of Orion's H\,II regions M\,42 and NGC\,2024. 
As explained before, 
the knots as we see them are most probably not single dust clouds, but 
cloud complexes with a mixture of several dust clouds with different 
individual temperatures. Therefore, the derived $L/M$ value is just an average 
and is expected to be larger for individual clouds, reaching the range of the 
Galactic compact H\,II regions. Hence, the warm knots are very good candidates
for containing not only moderate SF regions but also compact H\,II regions.

\subsection{Cold knots: Giant molecular cloud complexes}
The cold knots of SED type I 
represent giant molecular cloud complexes which contain at least a few
star forming regions.
The evidence comes from 
(1) the spatial coincidence with both the CO and H\,I detections and (2) 
the $L/M$ ratio 
(see Table \ref{l_m_tab})
which is higher 
for the cold knots
than for the rest of M\,31 tracing the ISRF
as discussed in Sect. \ref{sec_l_m}.   
For comparison, the $L/M$ values for the Chameleon clouds which exhibit 
only low mass star formation are clearly 
lower and even below the average of the Milky Way. 
Thus, at least a moderate SF also takes place in the cold knots of M\,31.

This picture might be supported by the distribution of the
thermal radio continuum as tracer for ionized gas. A comparison has been 
performed with figures 4 and 5 of Beck et al. (\cite{Beck+98}), showing the radio continuum 
at 20\,cm.
Except for knot 23, for all cold knots in the overlapping part radio continuum 
radiation is present, though only weak for knots 15, 17, and 37. Unfortunately
in M\,31 the synchroton radiation is very strong and still dominates the 
radio continuum at 20\,cm (Hoernes et al. \cite{Hoe+98}).
It will therefore be 
interesting to compare the FIR knots with the new 6\,cm Effelsberg survey
(Berkhuijsen et al., in prep.) and a corresponding thermal radio map resulting 
from the decomposition of the radio continuum.

\subsection{Medium knots: Mixture of warm and cold knots}

The knots of SED type II do probably represent a mixture containing 
both high mass star forming regions and cold molecular cloud complexes. 
Arguments in favour of this view come from the middle position 
in (1) cold dust temperatures and colours and (2) $L/M$ excess and 
(3) the average spatial coincidence with H\,II regions as well as CO and H\,I detections.

\subsection{The enigmatic knot \#\,23}
\label{shl23}
Knot \#\,23 northwest of M31's centre is by far the brightest 175\,$\mu$m 
region of M\,31. 
Despite its dominant 175\,$\mu$m appearance, the object does hardly 
show up in the IRAS 12 and 25\,$\mu$m HIRES maps, 
and gets only a little pronounced
in the 60\,$\mu$m map, especially when compared to the bright 
knots in the outer rings.
At 100\,$\mu$m it is clearly visible, but it is not the brightest region
of M\,31.
This object constitutes a real puzzle, since it is not prominent  
at any other tracers:
No radio continuum has been detected so far in this region (\cite{Braun90}; 
Beck et al. \cite{Beck+98}). Only diffuse and faint H\,I emission can be found
with no dominant feature at the position of the knot 
(Brinks \& Shane \cite{Bri84}).
The same yields for optical investigations. In Hodge (\cite{Hod80})
and Hodge (\cite{Hod81})
a lot of extinction is seen around the knot, but no specific dark cloud can 
be identified with the source. 
Whereas no CO is seen on the maps of Dame et al. (\cite{Dam+93}), new 
CO observations at higher resolution reveal a clumpy structure of small
and faint CO sources (\cite{Niet+00}).

With a temperature of 16.5\,K, knot \#\,23 belongs to the cold SED type I.
However, the value of $L/M = 450$ is still relatively high.
Between 100\,$\mu$m and 175\,$\mu$m the
intensity increases by a factor of 2, and we expect the maximum
to lie even beyond 175\,$\mu$m. In this case the temperature could be 
lower and the mass even higher, 
bringing $L/M$ closer to the 
value of about 150 for the central part powered only by the ISRF.
For this, the mass of the dust inside the measured aperture has to increase 
to at least $10^5$\,M$_\odot$. 
As the Jeans limit for
a single cloud with $R= 10^6$\,pc and $T=20$\,K can be estimated to
be about 500\,M$_\odot$, one will have 
problems to explain the stability of this complex without assuming SF.
Therefore, a pure heating from outside via the ISRF does not seem sufficient,
but further observations in the submm will clarify this point.
Additional clues will come from 
future observations of the thermal radio component of this region, 
and maybe even of the stellar population inside the knot. 
On the basis of the current data, also the 
possibility of a background object like a radio galaxy cannot be ruled out.

\begin{acknowledgement}
It is a pleasure for us to thank Rolf Chini, Kalevi Mattila and Viktor T\'oth 
for stimulating discussions, and the referee James Lequeux for his very helpful
comments improving this paper. 
The development and operation of ISOPHOT and the Postoperation Phase
are supported by funds of Deutsches Zentrum f\"ur Luft- und Raumfahrt 
(DLR, formerly DARA).
\end{acknowledgement}

\end{document}